\documentclass[]{elsart3p}

\usepackage{graphicx}
\usepackage{color}

\begin{document}

\journal{Physica A.}

\begin{frontmatter}

\title{Effective carrying capacity and analytical solution of a particular case of the Richards-like two-species population dynamics model}

\author[Filo]{Brenno Caetano Troca Cabella\corauthref{cor}}, 
\ead{brenno@usp.br}
\author[UFLA]{Fabiano Ribeiro} and 
\ead{fribeiro@dex.ufla.br}
\author[Filo,INCT]{Alexandre Souto Martinez},
\ead{asmartinez@ffclrp.usp.br}

\corauth[cor]{Corresponding author.}

\address[Filo]{ Faculdade de Filosofia, Ci\^encias e Letras de Ribeir\~ao Preto (FFCLRP), \\
             Universidade de S\~ao Paulo (USP) \\ 
             Avenida Bandeirantes, 3900 \\ 
             14040-901, Ribeir\~ao Preto, S\~ao Paulo, Brazil.}

\address[UFLA]{Departamento de Ci\^encias Exatas (DEX), \\
                Universidade Federal de Lavras (UFLA) \\
	           Caixa Postal 3037 \\
		     37200-000 Lavras, MG, Brazil.}

\address[INCT]{Instituto Nacional de Ci\^encia e Tecnologia em Sistemas Complexos}%

\begin{abstract}

We consider a generalized two-species population dynamic model and analytically solve it for the amensalism and commensalism ecological interactions. 
These two-species models can be simplified to a one-species model with a time dependent extrinsic growth factor.
With a one-species model with an effective carrying capacity one is able to retrieve the steady state solutions of the previous one-species model.
The equivalence obtained between the effective carrying capacity and the extrinsic growth factor is complete only for a particular case, the Gompertz model. 
Here we unveil  important aspects of sigmoid growth curves, which are relevant to growth processes and population dynamics.

\end{abstract}

\begin{keyword}
population dynamics (ecology) \sep 
generalized logarithmic and exponential functions \sep 
growth models \sep 
Richards model \sep
two-species model 
\PACS  89.75.-k \sep  
             87.23.-n \sep  
             87.23.Cc \sep 
\end{keyword}

\end{frontmatter}

\section{Introduction}
\label{intro}

Population dynamics models are useful when one tries to understand, describe or predict the behavior of a wide range of time dependent processes in several disciplines. 
To easily formulate and write the solutions of the population dynamics model, we present a one-parameter generalization of the logarithmic and exponential functions. 
This generalization has been first introduced in the context of non-extensive statistical mechanics~\cite{tsallis_1988,tsallis_qm,arruda_2008,martinez:2008b,Martinez:2009p1410}. 
The generalized logarithm or the so-called $\tilde{q}$-\textit{logarithm function} is defined as: 
\begin{equation}
\ln_{\tilde{q}}(x) = \lim_{\tilde{q}' \rightarrow \tilde{q}}\frac{x^{\tilde{q}'} -1 }{\tilde{q}'} = \int_1^x \frac{dt}{t^{1-\tilde{q}}}  \; .
\label{eq-ln-q}
\end{equation}
The natural logarithm function is retrieved for $\tilde{q} \rightarrow 0$.  
The inverse of the $\tilde{q}$-logarithm function  is the $\tilde{q}$-\textit{exponential function}
\begin{equation}
e_{\tilde{q}}(x) =   
\left\{ \begin{array}{ll}
\lim_{\tilde{q}^{'} \to \tilde{q}}   (1+ \tilde{q}^{'} x)^{1/\tilde{q}' } & ,\textrm{ if $\tilde{q}x > -1$} \\
0 & ,  \textrm{ otherwise} 
\end{array} \right. \; ,
\label{def-eq}
\end{equation} 
so that for $\tilde{q} =0$, one retrieves the usual exponential function.
The use of these functions is convenient since it has allowed us to find and simplify (using their properties) the solution of the models with time-dependent intrinsic and extrinsic growth rates.

The simplest way to deal with population growth is to consider \textit{one-species models}.
In these models, individuals do not explicitly interact with the external ones and, at time $t$, the number of individuals is $N(t) \ge 0$, with initial condition $N_0 \equiv N(0) > 0$. 
The parameters are the intrinsic growth rate $\kappa > 0$ and the environment carrying capacity $K = N(\infty) > 0$, which takes into account all possible interactions among individuals and resources~\cite{blanco_1993}. 
If $N(t)/K \ll 1$, a general model is the \textit{von Foerster et al.} model~\cite{vonfoerster}, where the per capita growth rate is $d \ln N/ dt = \kappa N^{\tilde{q}}$. 
Considering $\tau \equiv \kappa t \ge 0$ as a dimensionless time unity, the model solution is:  $N(\tau) = N_0 e_{-\tilde{q}}[\tau/(\tilde{q} T)]$, with $N_0 = (\tilde{q} T)^{-1/\tilde{q}}$ and produces a population size divergence at  a finite time $T$, obtained from the $\tilde{q}$-exponential $\tilde{q} \tau /(\tilde{q} T) > 1$~\cite{Strzalka:2009p2964}. 
As $\tilde{q} \to 0$, the exponential population growth (\textit{Malthus} model) $N(t) = N_0 e^{\kappa t}$, with divergence at an infinite time, is obtained.
The population size divergence can be dismantled considering a finite carrying capacity. 

The richness ecological community cannot be properly described only by the one-species models.
The interactions between species becomes better formalized  with the (Malthus-like) Lotka-Volterra equations~\cite{murray}, which only explains prey-predator behavior in its original formulation and presents stability problems.  
Besides the predation interaction, there are many other different kinds of interaction taking place between two biological species. 
For instance, if species negatively affects each other, when they occupy the same ecological niche and use the same resources, there is \textit{competition}.  
If species favor each other, there is \textit{mutualism} or \textit{symbiosis}.  
These ecological richness is partially appreciated in the competitive (Verhulst-like) Lotka-Volterra model studied in Ref.~\cite{ribeiro_2011_2}. 
This model presents complete analytical solutions. 
It has a non-trivial phase diagram and solves the stability problem of the Malthus-like two species model. 

Here, we consider a generalized two-species population dynamic model and analytically solve it for particular ecological interactions. 
These two-species models can be simplified to a one-species model, with a time dependent extrinsic growth factor.
With a one-species model with an effective carrying capacity, one is able to retrieve the steady state solutions of the previous one-species model. 
The equivalence obtained between the effective carrying capacity and the extrinsic growth factor is complete only for the Gompertz model. 

Our presentation is organized as follows:
In the Sec.~\ref{gen1}, the Richards model is presented in terms of the generalized functions, which has the Verhulst and Gompertz models as particular cases.
We show that the steady state solution of the Richards model with extrinsic growth rate is the same for the Richards model without extrinsic growth rate, but with a modified carrying capacity. 
Also, for a particular case of the Richards model ($\tilde{q} \rightarrow 0$), that is the Gompertz model, not only the steady state is the same but also the whole system evolution (transient).
In the Sec.~\ref{gen2}, we present a generalized model of two interacting species. 
We show that the interaction between species can also be interpreted as an extrinsic growth factor. 
This allows one to represent a two-species system by a one-species model with a modified carrying capacity.
In Sec.~\ref{conclusion}, we present our final remarks. 

\section{Generalized one-species model}
\label{gen1}

The growth of individual organisms~\cite{laird65}, tumors~\cite{bajzer96} and other biological systems~\cite{zwietering90} are well described by mathematical models considering a finite carrying capacity $K$, which lead to sigmoid growth curves~\cite{boyce_diprima,murray,Keshet}.  
In this context, it is convenient to express the population size $N$ with respect to its equilibrium value, i.e. $p=N/K$. 
The \textit{Richards} model~\cite{richards_1959}
\begin{equation}
\frac{d }{d \tau} \ln p  =  \ln_{\tilde{q}} p,
\end{equation}
binds the  \textit{Gompertz} ($\tilde{q} = 0$, $d \ln p/d \tau = \ln p$) and the \textit{Verhulst} ($\tilde{q} = 1$, $d \ln p/d  \tau = 1 - p$) models through the $\tilde{q}$ parameter, which microscopic interpretation is given in Refs.~\cite{martinez:2008b,Mombach:2002p1065,donofrio}.
This parameter is related to the range of interaction between the individuals that compose the population and the fractal pattern of the population structure.
This generalization is corroborated by several different approaches in terms of Tsallis statistics \cite{Strzalka:2008p2511} and logistic equations of arbitrary order \cite{grabowski:2010p3081} for the classical Verhulst and Malthus models.
The solution of the Richards model is: 
\begin{equation}
 p(\tau)=\frac{1}{e_{\tilde{q}}\{\ln_{\tilde{q}}[1/p_0]e^{-\tau}\}},
\end{equation}
whose initial condition is $p_0 \equiv p(0)$ and the  steady state solution is $p^* = p(\infty) = 1$  \cite{brenno_2011}. 

Adding an extrinsic growth rate $\epsilon$ to the Richards' model, one has the so called \textit{Richards-Schaefer} model and respective solution: 
\begin{eqnarray}
\frac{d \ln p}{d\tau} & = & - \ln_{\tilde{q}}p + \epsilon 
\label{steady_state_ii} \\
p(\tau)                      & = & \frac{e_{\tilde{q}}(\epsilon)}{e_{\tilde{q}}\{\ln_{\tilde{q}}[e_{\tilde{q}}(\epsilon)/p_0]e^{-[1+\tilde{q}\epsilon]\tau}\}} \; ,
\label{transient_ii}
\end{eqnarray}
with steady state $p(\infty) \equiv  p^*=e_{\tilde{q}}(\epsilon)$, which behaves as an order parameter. 
The Richards' model solution is retrieved for $\epsilon = 0$. 
The parameter $\epsilon$ plays the role of an external factor that removes or insert individuals in the population. 
For instance, it can be an interaction factor between different species, or the effect of the cancer treatment when one deals with tumor cells population. 
Survival of the species is obtained when $\tilde{q} \epsilon > -1$, and extinction occurs otherwise.
The critical value $\epsilon^{(c)} = -1/\tilde{q}$ separates the extinction from the survival phase, which behaves as $p^* \sim [\epsilon - \epsilon^{(c)}]^{1/\tilde{q}}$ near the critical point.
This transition occurs for $\tilde{q} \ne 0$, so that only the Gompertz model does not present the extinction-survival transition. 

From the analytical solution of Eq.~(\ref{transient_ii}), we show that the insertion ($\epsilon > 0$) or removal ($\epsilon < 0$) of individuals from the population  modifies the steady state solution. 
Instead of considering the population size with respect to the medium (bare) carrying capacity, we propose to consider the population size with respect to a rescaled carrying capacity $K'= e_{\tilde{q}}(\epsilon)K$. 
It means that the $N$ population individuals  live in an environmental with  an effective carrying capacity $K'$, but now without insertion or removal of individuals.  
The population size with respect to the new equilibrium value is $p'= N/K'= p/e_{\tilde{q}}(\epsilon)$ and one has the Richards' model and its respective solution:  
\begin{eqnarray}
\frac{d \ln p' }{d\tau} & = & -\ln_{\tilde{q}}\left[ p'\right] \; , 
\label{steady_state_iii} \\
p'(\tau) & = & \frac{p(\tau)}{e_{\tilde{q}}(\epsilon)} =\frac{1}{e_{\tilde{q}}[\ln_{\tilde{q}}(1/p_0')e^{-\tau}]} \ ;
\label{transient3}
\end{eqnarray}
with steady state: $p'(\infty) = (p')^* = 1$, which of course leads to: $p^*=e_{\tilde{q}}(\epsilon)$ in the original variable. 
Observe that the Eq.~(\ref{steady_state_iii}) is valid only for the survival phase, i.e. $\tilde{q} \epsilon > -1$, otherwise $K'$ vanishes.

\begin{figure}[htbp]
	\centering
		\includegraphics[width=.95\columnwidth]{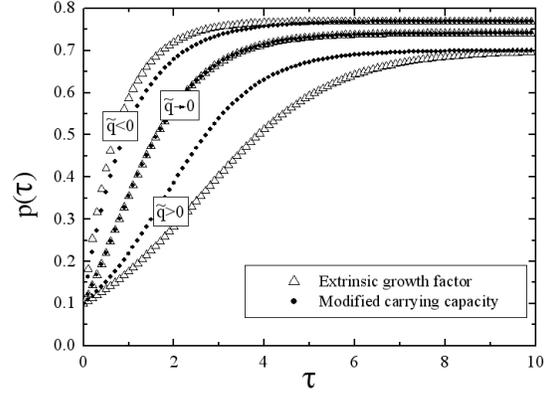}
	\caption{Plot of population $p(\tau)$ [Eqs.~\ref{transient_ii} and~\ref{transient3}] as a function of $\tau$, with: $p_0=0.1$, $\epsilon=-0.3$, $\tilde{q}=-1$, $\tilde{q} \rightarrow 0$  (Gompertz model) and $\tilde{q}=1$  (Verhulst model). 
The full equivalence of the models with extrinsic growth factor and modified carrying capacity is only obtained for the steady state solutions. 
Nevertheless, for $\tilde{q} \rightarrow 0$, the transient solution is also equivalent.}
\label{fig:equivalence}
\end{figure}

Comparing the solutions of Eqs.~(\ref{transient_ii}) and~(\ref{transient3}), one sees that the considered models present the same steady state solutions.  
Nevertheless, they have different  transient behavior, except  for Gompertz model ($\tilde{q}\rightarrow 0$), where the equivalence between the original and rescaled models is complete.
The plots of Fig.~\ref{fig:equivalence} illustrate this behavior. 
In fact, this property comes from the argument of the exponential function: while in Eq,~(\ref{transient_ii}) this argument is $-\tau$ , in Eq.~(\ref{transient3}) the argument is $-(1+\tilde{q}\epsilon)\tau$. 
Thus the evolution of the two systems become identical only if $\tilde{q} \epsilon =0$. As we are dealing with $\epsilon \ne 0$, the evolution of the two model is the same only when $\tilde{q} \rightarrow 0$ i.e. the Gompertz model.

\section{Generalized two-species model}
\label{gen2}

Here, we introduce a generalization of the Verhulst-like two species model, considered in Ref.~\cite{ribeiro_2011_2}, which we call the \textit{Richards-like two-species model}. 
We consider the Richards term instead of the Malthus one in the original Lotka-Volterra equations. 
In this type of model, the two species interact according to the Lotka-Volterra equation, however the species also suffer from inter-species competition.  
As in Ref.~\cite{ribeiro_2011_2}, we let the interaction parameter vary from negative to positive values to cover all possible ecological regimes. 
The proposed model is written as: 
\begin{eqnarray}
 \frac{d p_1}{d\tau} &=&          p_1[  \ln_{\tilde{q}_1} p_1 + \epsilon_1 p_2]   \label{generalized-model_1} \\
 \frac{d p_2}{d\tau} &=&  \rho p_2[ \ln_{\tilde{q}_2} p_2 + \epsilon_2 p_1]    \label{generalized-model_2} \; , 
\end{eqnarray}
where $p_i = N_i/K_i \ge 0$, for $i=1,2$, where $N_i \ge 0$, $\kappa_i$ and $K_i > 0$ are the number of individuals (size), net reproductive rate, and the carrying capacity of species $i$,  respectively. 
The carrying capacity $K_1$ represents the restriction on resources that comes from any kind of external factors that do not have to do with species 2 and 
similarly for $K_2$. 
Time is measured with respect to the net reproductive rate of species 1, $\tau = \kappa_1 t \ge 0$. 
The scaled time is positive since we take the initial condition at $t_0 = 0$. 
Moreover, the two net reproductive rates form a single parameter  $\rho = \kappa_2/\kappa_1 > 0$, fixing a second time scale:  $\tau' \equiv \rho \tau = \kappa_2 t$. 
In the right side of Eq.~(\ref{generalized-model_1}), the term $p_1  \ln_{\tilde{q}_1} p_1$ represents the competition between individuals of the same species (intraspecific competition) and $\epsilon_1 p_1 p_2$ represents the interaction between individuals of different species (interspecific interaction)~\cite{murray,Keshet}.
The right side of Eq.~(\ref{generalized-model_2}) behaves similarly.
The non-dimensional population interaction parameters are   $\epsilon_1 $ and  $\epsilon_2$, which are not restricted and may represent  different ecological interactions. 
Contrary to $\rho$, which has no major relevance to this model (since we consider only $\rho>0$), the product $\epsilon_1\epsilon_2$ plays an important role, so that $\epsilon_1 \epsilon_2 < 0$ means predation;  $\epsilon_1\epsilon_2 = 0$ means commensalism, amensalism, or neutralism; and  $\epsilon_1 \epsilon_2 > 0$ means either mutualism or competition. 
For $\tilde{q}_1 = \tilde{q}_2 = 1$, one retrieves the model of  Ref.~\cite{ribeiro_2011_2}. 

Here, we restricted ourselves to the particular case when $\epsilon_1 \epsilon_2 = 0$. 
Thus,  only one species is affected by the other or they do not interact at all.  
Closed analytical solutions can be found in this particular case.  
Consider that the individuals of species $1$ is unaffected by species $2$. 
In this case, species $1$ is described by Eqs.~\ref{steady_state_ii} and~\ref{transient_ii}, so that:
\begin{eqnarray}
\frac{d \ln p_1}{d\tau} & = & - \ln_{\tilde{q}_1}p_1
\label{steady_state_i_a} \\
p_1(\tau) & = & \frac{1}{e_{\tilde{q}_1}[\ln_{\tilde{q}_1}(p_{1,0}^{-1})e^{-\tau}]} \; ,
\label{transient_i_a}
\end{eqnarray}
where the initial condition is $p_{1,0} = p_1(0)$ and steady state: $p_1(\infty) = p_1^*=1$. 
The individuals of species $2$ are affected by species $1$ through an extrinsic growth factor $\epsilon'_2(\tau)$, and one writes the equation for species $2$ as:
\begin{eqnarray}
\frac{d \ln p_2}{d\tau} & = & -\rho \left[ \ln_{\tilde{q}_2}p_2-\epsilon_2'(\tau) \right] \; 
\label{steady_state_i}
\end{eqnarray}
where,
\begin{equation}
\epsilon'_2(\tau)=\epsilon_2 p_1(\tau)=\frac{\epsilon_2}{e_{\tilde{q}_1}[\ln_{\tilde{q}_1}\left(p_{10}^{-1}\right) e^{-\tau}]} \; .
\label{epsilon2l}
\end{equation}
The solution of Eq.~(\ref{steady_state_i}) has been obtained in Ref.~\cite{brenno_2011}:
\begin{eqnarray}
p_2(\tau) & = & \frac{e_{\tilde{q}_2}[\epsilon_2'(\tau)]}
                     {e_{\tilde{q}_2}\left\{
                      \ln_{\tilde{q}_2} \left\{\frac{e_{\tilde{q}_2} [\epsilon_2'(0)]}{p_{2,0}} \right\} 
                      \frac{e_{\tilde{q}_2}[\epsilon_2'(0)]}
                              {e_{\tilde{q}_2}[\epsilon_2'(\tau)]} 
                             e^{- \left[1 + \tilde{q}_2 \overline{\epsilon}_2'(\tau) \right] \rho \tau}
                     \right\}} \; ,
\label{transient_i}
\end{eqnarray}
where the initial condition is $p_{2,0} = p_2(0) $ and steady state: $p_2(\infty) = p_2^*=e_{\tilde{q}_2}(\epsilon_2)$. 
In Eq.~(\ref{transient_i}), $\overline{\epsilon}_2'(\tau)  =  (1/\tau) \int_0^{\tau} d \tau' \epsilon_2'(\tau')$ is the  mean value of $\epsilon_2'(\tau)$ up to $\tau$.

\begin{figure}[htbp]
	\centering
		\includegraphics[width=.95\columnwidth]{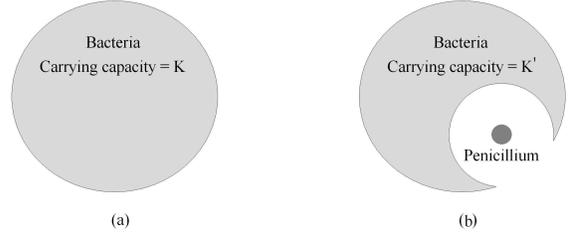}
	\caption{Representation of the amensalism ecological interaction between the bread mold and bacteria. 
	               A two-species model can be simplified to an one species model with a modified carrying capacity (gray area). 
	               {\bf (a)} Bacteria occupying all the available space with carrying capacity $K$. 
	               {\bf (b)} Restricted bacteria growth due to the interaction with bread mold, carrying capacity $K'<K$.}
	\label{fig:penicillium}
\end{figure}

For $\epsilon_2 < 0$ ($\epsilon_2 > 0$), one has amensalism (commensalism), so that species $1$ adversely (positively) affects species $2$. 
The bread mold penicillium is a common example of amensalism. 
The analytical solution for the amensalism and commensalism regimes is obtained considering the time dependent extrinsic growth factor as a function of species $1$, according to Eq. (\ref{epsilon2l}). 
Penicillium secretes penicillin, a chemical that kills bacteria. 
Consider the two-species model of Eqs.~(\ref{steady_state_i_a}) and~(\ref{steady_state_i}), with $\epsilon_2<0$, the interaction term $\epsilon_2'$ can be interpreted as an extrinsic growth rate of a one-species model. 
As we have seen, an extrinsic factor can be incorporated into the carrying capacity. 
In this way, a two species model can be interpretated as a one-species model, with modified carrying capacity, as depicted in Fig \ref{fig:penicillium}.  

\begin{figure}[htbp]
	\centering
      \includegraphics[width=.95\columnwidth]{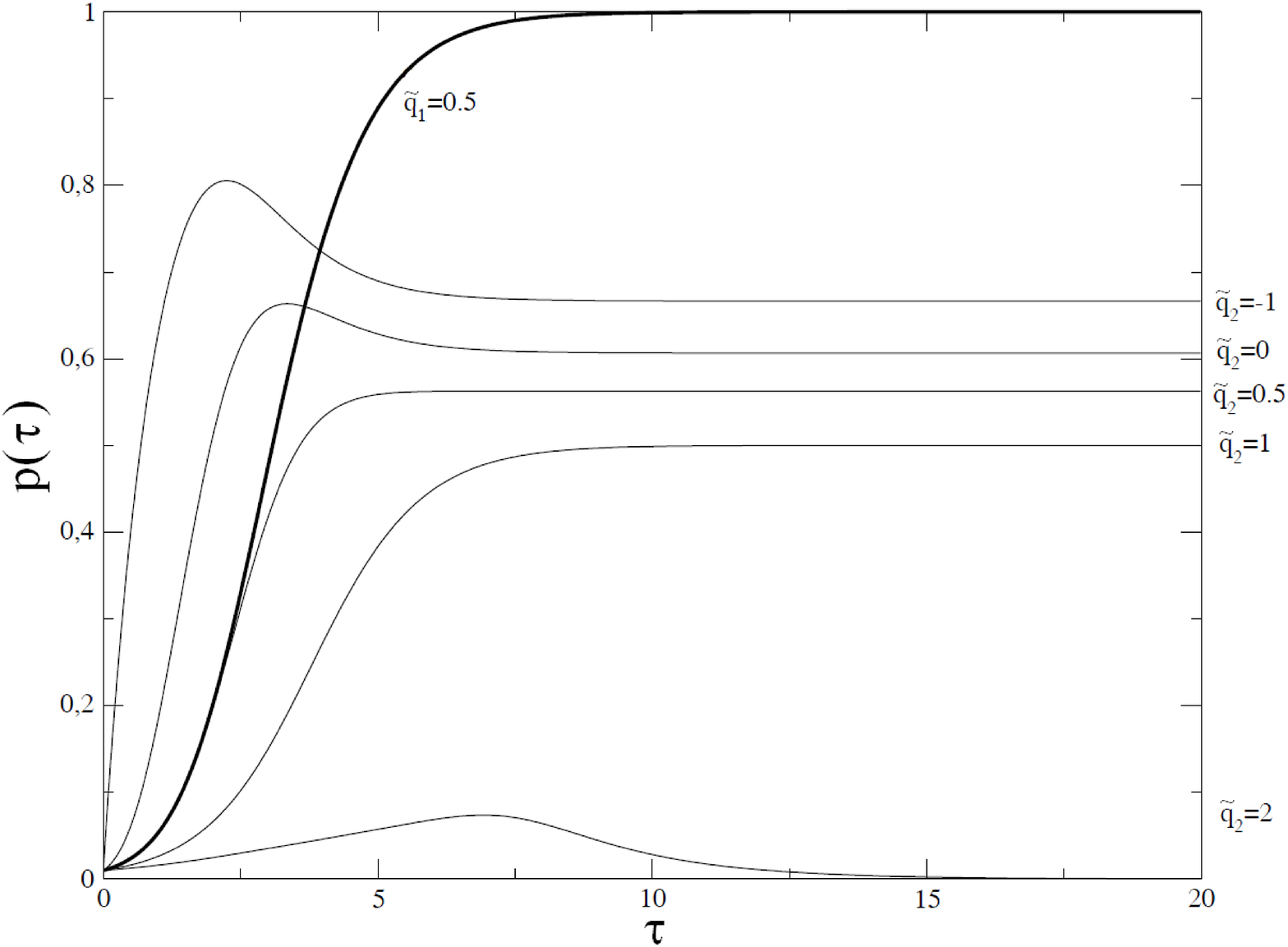} {\bf (a)} \\
	\includegraphics[width=.95\columnwidth]{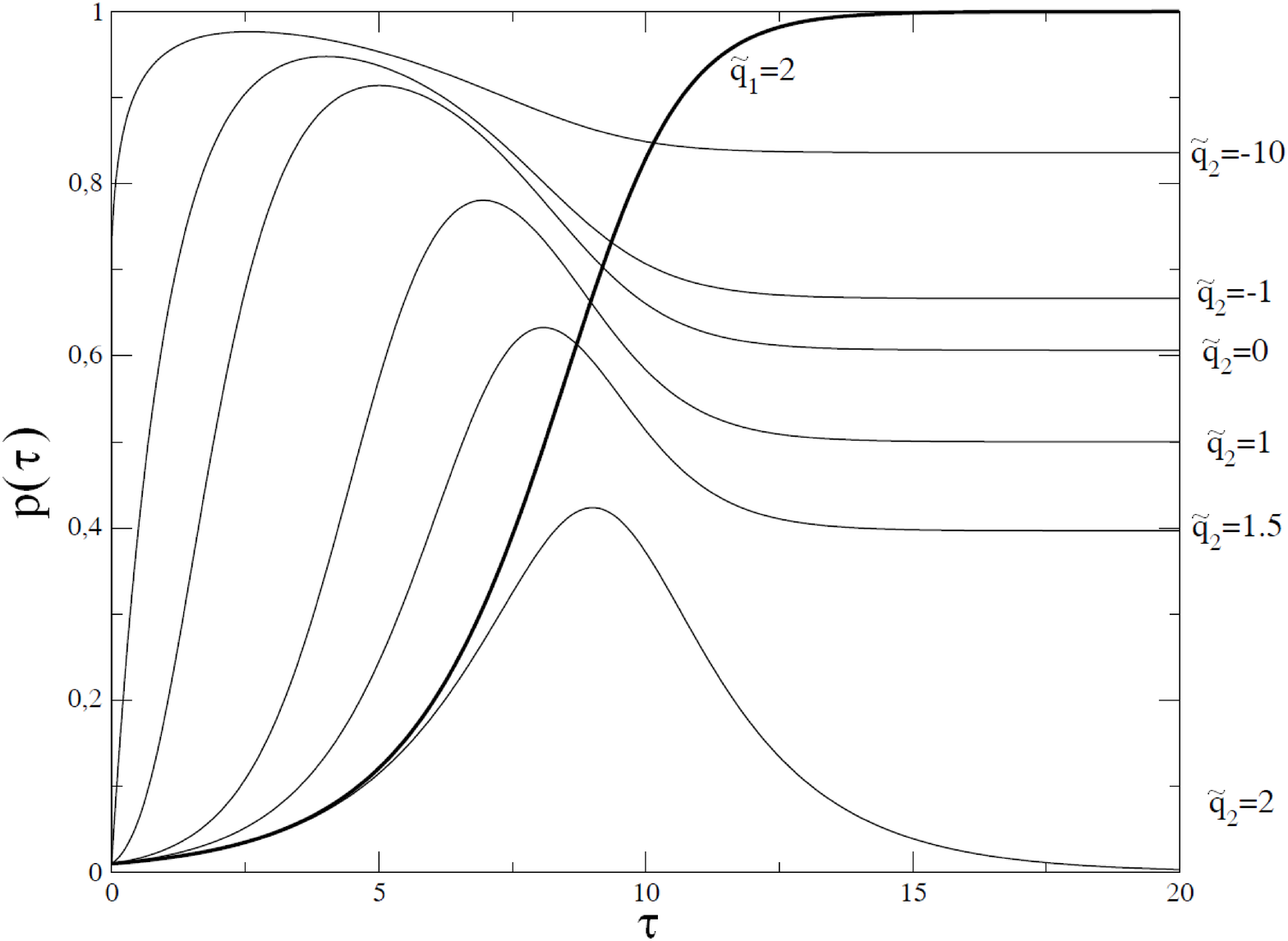} {\bf (b)} 
\caption{For several values of $\tilde{q}_2$, plots of $p_1(\tau)$ (thick lines), given by Eq.~\ref{transient_i_a} and of $p_2(\tau)$ (thin lines), given by Eq.~\ref{transient_i}, where the integral of Eq.~\ref{transient_i} has been calculated numerically. 
In both cases $\rho = 1$, $\epsilon_1 = 0$ and $\epsilon_2 = 1/2$ :  {\bf (a)} $\tilde{q}_1 = 1/2$ and  {\bf (b)}  $\tilde{q}_1 = 2$. 
Notice that if $\epsilon_2 \tilde{q}_2 < -1$, $p_2^* = 0$.}
	\label{amensa-out-eq}
\end{figure}

Species 2 evolves as a time dependent one-species (Richards-Schaefer) model, whose solution has been obtained in Ref.~\cite{brenno_2011} and extinction occurs for $\epsilon_2'(\tau) \tilde{q}_2 < -1$. 
The evolution of both species is presented in Fig.~(\ref{amensa-out-eq}) for two different values of $\tilde{q}_1$.
The time transient depends on species 1, through the effective extrinsic factor $\epsilon_2'(\tau)$. 
For $\tilde{q}\ne0$, there is a dependence on the mean value $\overline{\epsilon}_2'(\tau)$, see Eq.~(\ref{transient_i}).
Considering the Gompertz model ($\tilde{q}_2 \rightarrow 0$), the solution can be simplified to: 
\begin{eqnarray}
p_2(\tau) = e^{\epsilon_2'(\tau)}\left[\frac{p_{2,0}}{e^{\epsilon_2'\left(0\right)}}\right]^{e^{[\epsilon_2'(0) - \epsilon'_2(\tau)-\rho\tau]}}.
\label{p2taugom}
\end{eqnarray}

\section{Conclusion}
\label{conclusion}

We have proposed a generalized two-species population model and have found its analytical solution for the commensalism and amensalism regimes. 
From this result, we are able to show that the solution of the two-species model is equivalent when one considers one-species model with a time dependent extrinsic growth factor.
Also, the extrinsic growth factor can be incorporated into the carrying capacity and the full equivalence, i.e. steady state and transient solution, is retrieved for the Gompertz model. 
Otherwise, the equivalence persists only for the steady state solutions.
This feature reveals an important aspect of sigmoid growth curves.
Considering two well known growth models, Verhulst  and Gompertz , besides presenting a similar sigmoid behavior, they are differently affected by an extrinsic growth factor. 
For the Gompertz model, the extrinsic growth factor can be fully incorporated into the carrying capacity; i.e. removing individuals from the population has the same effect if one considers that this population grows in a more limited environment. Any kind of external influence can be seen as a modification in the environment limitations.
Although the extrinsic factor can also be incorporated in the Verhulst model, it correctly describe only the steady state solution, the equivalence is lost when one considers the transient solution.

\section*{Acknowledgements}

B. C. T. C. acknowledges support from CAPES.
F. R. acknowledges support from CNPq (151057/2009-5). 
A. S. M. acknowledges the Brazilian agency CNPq (305738/2010-0 and 476722/2010-1) for support. 


\end{document}